\documentstyle[epsfig]{aipproc}

\catcode`\ä = \active \catcode`\ö = \active \catcode`\ü = \active \catcode`\Ä = \active
\catcode`\Ö = \active \catcode`\Ü = \active \catcode`\ß = \active \catcode`\é = \active
\catcode`\è = \active \catcode`\ë = \active \catcode`\ô = \active \catcode`\ê = \active
\catcode`\ø = \active \catcode`\ò = \active
\defä{\"a} \defö{\"o} \defü{\"u} \defÄ{\"A} \defÖ{\"O} \defÜ{\"U} \defß{\ss} \defé{\'{e}}
\defè{\`{e}} \defë{\"{e}} \defô{\^{o}} \defê{\^{e}} \defø{\o} \defò{\`{o}}

\begin{document}

\title{Collective enhancement and suppression in Bose-Einstein condensates}

\author{Wolfgang Ketterle, Ananth P. Chikkatur, and Chandra Raman}
\address{Department of Physics and Research Laboratory of Electronics, \\
Massachusetts Institute of Technology, Cambridge, MA 02139, USA}

\maketitle

\begin{abstract}
The coherent and collective nature of Bose-Einstein condensate
can enhance or suppress physical processes.  Bosonic stimulation
enhances scattering in already occupied states which leads to atom
amplification, and the suppression of dissipation leads to
superfluidity. In this paper, we review several experiments where
suppression and enhancement have been observed and discuss the
common roots of and differences between these phenomena.
\end{abstract}

When a gas of bosonic atoms is cooled below the transition
temperature of Bose-Einstein condensation, it profoundly changes
its properties.  The appearance of a macroscopically occupied
quantum state leads to a variety of new phenomena which set
quantum fluids apart from all other substances.  Fritz London even
called them the fourth state of matter~\cite{lond261}.

Many of the key concepts in quantum fluids were derived from
studying the weakly interacting Bose gas, for which rigorous
theoretical treatments were possible \cite{huan87,bogo47}. In
1995, with the discovery of BEC in a dilute gas of  alkali atoms
\cite{ande95,davi95evap,brad95bec}, it became possible to study
such a system experimentally . The theoretical framework connects
the observed equilibrium and dynamic properties to the presence
of long-range order and low-lying collective excitations
\cite{dalf99rmp}.

Many special properties of Bose condensates involve the
suppression or enhancement of physical processes.  Our recent
experiments include the suppression and enhancement of elastic
collisions of impurity  atoms \cite{chik00}, the suppression of
dissipation due to superfluidity \cite{rama99,onof00sup}, and the
suppression \cite{stam99phon} and enhancement of light scattering
\cite{inou99super}. Bosonically enhanced Rayleigh scattering was
used to amplify either atoms \cite{inou99mwa} or light
\cite{inou00slow} in a condensate dressed by laser light. We
review these experiments and discuss the properties of the Bose
condensate which lead to enhancement and suppression.

\section{Scattering of light and massive particles}
\label{sec:scattering}

Before we discuss light scattering and collisions in a BEC, we want to derive some simple general
expressions based on Fermi's golden rule which will be useful to see the similarities and
differences between the different processes.

When a condensate scatters a photon or material particle, the scattering is described by the
Hamiltonian
\begin{equation}\label{eq:hamil}
{\cal H^\prime} = C \sum_{k,l,m,n} \hat{c}^\dagger_{l}
  \hat{a}^\dagger_n \hat{c}_k \hat{a}_m \delta_{l+n-k-m}
\end{equation}
Here $\hat{c}_k$ ($\hat{c}^\dagger_k$) is the destruction (creation) operator for the scattered
particles, and $\hat{a}_k$ ($\hat{a}^\dagger_k$) is the destruction (creation) operator for atomic
plane waves of wavevector $\bf k$.  The strength of the coupling is parametrized by the coefficient
$C$.

We consider the scattering process where a system with $N_0$
atoms in the condensate ground state and $N_q$ quasi-particles
with wavevector $\bf q$ scatters particles with incident momentum
$\bf k$ into a state with momentum $\bf k - q$.  The initial and
final states are

\begin{eqnarray}\label{eq:states}
|i\rangle &=&  |n_k, n_{k-q} ; N_0, N_q\rangle  \nonumber \\
|f\rangle &=&  |n_k-1, n_{k-q} +1 ; N_0-1, N_q+1\rangle
\end{eqnarray}
respectively, where $n_k$ denotes the population of scattering
particles with wavevector $\bf k$\footnote{This choice of final
states implies that we neglect scattering between quasi-particles
and consider only processes involving the macroscopically
occupied zero-momentum state of the condensate. Formally, we
replace the Hamiltonian (Eq.\ \ref{eq:hamil}) by $C \sum_{k,q}
(\hat{c}^\dagger_{k-q}\hat{a}^\dagger_q \hat{c}_k \hat{a}_0 +
\hat{c}^\dagger_{k-q} \hat{a}^\dagger_0 \hat{c}_k
\hat{a}_{-q})$.}. It should be emphasized that, due to the
interatomic interactions, the quasi-particles with occupation
$N_q$ are not the plane waves created by the operator
$\hat{a}^\dagger_q$, but the quanta of collective excitations with
wavevector $\bf q$.

The square of the matrix element $M_1$ between the initial and final state is
\begin{eqnarray}\label{eq:matrix-el1}
  &&|M_1|^2 = |\langle f|{\cal H^\prime}|i \rangle|^2 \nonumber\\
  &&= |C|^2 |\langle N_0=N-1,N_q=1|\hat{\rho}^\dagger(\vec{q})|N_0=N,N_q=0
  \rangle|^2 (N_q+1) (n_{k-q} +1) n_k
\end{eqnarray}
where $\hat{\rho}({\bf q}) = \sum_m \hat{a}^\dagger_{m+q} \hat{a}_m$ is the Fourier transform of
the atomic density operator at wavevector $\vec{q}$.

The  static structure factor of the condensate is $S(q) = \langle g | \hat{\rho}({\bf q})
\hat{\rho}^\dagger({\bf q}) | g \rangle/N$ where $| g \rangle=|N_0=N, N_q=0 \rangle$ is the BEC
ground state. We then obtain for the scattering matrix element $M_1$

\begin{equation}\label{eq:matrix-el2}
  |M_1|^2 = |C|^2 S(q) (N_q+1) (n_{k-q} +1) N_0 n_k
\end{equation}

The scattering rate $W_1$ for the process $|n_k, n_{k-q} ; N_0, N_q\rangle \rightarrow | n_k-1,
n_{k-q} +1 ; N_0-1, N_q+1 \rangle$ follows from Fermi's golden rule as
\begin{equation}\label{eq:rate}
  W_1 = \frac{2 \pi}{\hbar} |M_1|^2  \delta(E_k - E_{k-q} - \hbar \omega_q^B)
\end{equation}
where $E_k$ is the energy of the incident particle with momentum
$\bf k$, and $\hbar \omega_q^B$ is the energy of quasi-particles
with momentum $\bf q$ (which we will later obtain from Bogoliubov
theory). To obtain the net scattering rate, one has to include
the reverse process  $|n_k, n_{k-q} ; N_0, N_q\rangle \rightarrow
|n_k+1, n_{k-q}-1; N_0+1, N_q-1 \rangle$ by which atoms scatter
{\it back} into the condensate. The square of the matrix element
$M_2$ for this process is $|C|^2 S(q) N_{q}n_{k-q} (
N_{0}+1)(n_{k}+1)$.
 The {\it net} rate $W_+$ of scattering
atoms from the condensate into the quasi-particle mode $\bf q$ is
the difference of the two partial rates $W_+=W_1-W_2$. Assuming
$N_0\gg 1$ (i.e. $N_0+1 \approx N_0)$, we obtain for the net rate
\begin{equation}\label{eq:net-rate}
  W_+ =\frac{2 \pi}{\hbar} |C|^2 S(q) N_0 [n_k (N_q + n_{k-q} +1)-N_q n_{k-q}] \delta(E_k
- E_{k-q} - \hbar \omega_q^B)
\end{equation}

For large $n_k$ (e.g.\ a laser beam illuminating the condensate)
the dominant bosonic stimulation term $(N_q + n_{k-q} +1)$ is
approximately $(\tilde{N}+1)$ with $\tilde{N}=\max(N_q, n_{k-q})$.
This illustrates that there is no bosonic stimulation of the net
rate by the {\it least} populated final state.  With the dynamic
structure factor $S({\bf q},\omega)=S(q) \delta(\omega
-\omega_q^B)$ Eq.\ \ref{eq:net-rate} simplifies to
\begin{equation}\label{eq:net-rate2}
  W_+/N_0=\frac{2 \pi}{\hbar^2} |C|^2 n_k \: S({\bf q},\omega) (N_q + n_{k-q} +1)
\end{equation}

The rate $W_+$ in Eq.\ \ref{eq:net-rate} is the rate for the
Stokes process where $E_k > E_{k-q}$. Momentum transfer $\bf q$
to the condensate is also possible as an anti-Stokes process
where a quasi-particle with momentum $\bf -q$ is scattered into
the condensate, and the scattered particle {\it gains} energy. The
net rate $W_-$ for this process is obtained in an analogous way as
\begin{equation}\label{eq:anti-stokes-rate}
  W_- = \frac{2 \pi}{\hbar} |C|^2 S(q) N_0 [n_k (N_{-q} - n_{k-q})-(N_{-q}+1) n_{k-q}]
   \delta(E_k - E_{k-q} + \hbar \omega_q^B).
\end{equation}

The net scattering rates in Eqs.\ \ref{eq:net-rate} and
\ref{eq:anti-stokes-rate} are the product of three terms.
\begin{itemize}
  \item The static structure factor $S(q)$ represents the squared matrix element for the condensate to
  absorb momentum $\bf q$.
  \item The delta function denotes the density of final states.
  \item The bosonic stimulation term represents stimulation by the occupation in the final state
  either of the scattering particles or the condensate.
\end{itemize}

The interplay of these three terms is responsible for the suppression and enhancement of physical
processes in a condensate.  The properties of the condensate as an intriguing many-body system are
reflected in the structure factor.  In section \ref{sec:Bragg_spect}, we discuss its measurement
through stimulated light scattering.  The density of states is responsible for superfluidity
because it vanishes for initial velocities of the incident particles which are smaller than the
Landau critical velocity (Sections \ref{sec:collisions} and \ref{sec:macroscopic_flow}). Finally,
bosonic stimulation by the occupancy $N_q$ of final states was responsible for superradiance,
matter wave amplification and optical amplification in a condensate (Section
\ref{sec:amp_light_atoms}).

\section{Determination of the static structure factor by Bragg spectroscopy}
\label{sec:Bragg_spect}

The BEC matrix element or the structure factor in Eqs.
\ref{eq:net-rate} and \ref{eq:anti-stokes-rate} can be directly
determined experimentally by stimulated light scattering.  The
density of states (the $\delta$ function in Eq.
\ref{eq:net-rate}) does not restrict the scattering since the
photon energy is much higher than the quasi-particle energy.  As a
result, photons can be scattered into the full solid angle and
provide the necessary recoil energy of the atom by a small change
($\approx 10^{-9}$) in the frequency of the scattered photon.

\begin{figure}
    \epsfxsize=70mm
 \centerline{\epsfbox{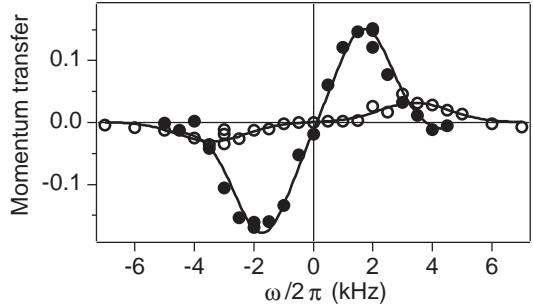}}
    \caption{ Bragg scattering of phonons and of free particles.
    Momentum transfer per particle,
    in units of $\hbar q$, is shown vs.\ the frequency difference
    $\omega / 2 \pi$ between the two Bragg beams.  The two spectra
    were taken at different atomic densities.  The open symbols
    represent the phonon excitation spectrum for a trapped condensate
    (at a chemical potential $\mu / h = 9.2$ kHz, much larger than
    the free recoil shift of $\approx 1.4$ kHz). Closed symbols show
    the free-particle response of a twenty-three times more dilute
    (ballistically expanded) cloud. Lines are fits to the difference
    of two Gaussian line shapes representing excitation in the
    forward and backward directions. See Refs.\
    \protect\cite{stam99phon,stamp00leshouches} for more details.
    Figure is taken from Ref.\ \protect\cite{stam99phon}. }
    \label{fig:phononspectra}
\end{figure}

When the condensate is illuminated by two strong laser beams with
wavevectors $\bf k$ and $\bf k-q$ and a difference frequency
$\omega$, the rate $W$ of transferring photons from one beam to
the other is
\begin{equation}
W/N_0=(W_+ + W_-)/N_0= 2 \pi \omega_R^2 S(q) (\delta(\omega
-\omega_q^B)-\delta(\omega +\omega_q^B))
\label{eq:total-stim-rate}
\end{equation}
where the two-photon Rabi frequency $\omega_R$ is given by
$\omega_R^2= |C|^2 n_k n_{k-q}/\hbar^2$. When $\omega$ is scanned
the ``spectrum'' of a condensate consists of two peaks at $\pm
\omega_q^B$ (Fig. \ref{fig:phononspectra}). The strength (or
integral) of each peak corresponds to $S(q)$.  We refer to this
method as Bragg spectroscopy since the basic process is Bragg
scattering of atoms from an optical standing wave
\cite{sten99brag}.

We observed that the scattering rate was strongly suppressed when
the momentum transfer $\hbar {\bf q}$ became smaller than the
speed of sound $c$ in the condensate, i.e.,\ when the light
scattering excited a phonon and not a free particle.  These
observations are in agreement with the Bogoliubov theory, which
obtains the elementary excitations or Bogoliubov quasiparticles as
eigenstates of the Hamiltonian (Eq. \ref{eq:hamil}).  A
quasi-particle with wavevector $q$ is annihilated by the
Bogoliubov operator $\hat{b}_{q} = u_q \hat{a}^{\dagger}_q + v_q
\hat{a}_{-q}$, where $u_{q} = \cosh \phi_{q}$, $v_{q} = \sinh
\phi_{q}$ and $\tanh 2 \phi_{q} = \mu / (\hbar \omega_{q}^{0} +
\mu)$.  Its energy is given by
\begin{equation}
\hbar \omega_q^{B} = \sqrt{ \hbar \omega^{0}_{q} (\hbar \omega^{0}_{q} + 2 \mu) },
\label{eq:Bogoliubov-energies}
\end{equation}
where $\hbar \omega^{0}_{q} = \hbar^2 q^2 / 2 m$ and $\mu$ is condensate's chemical potential.
Expressing  $\hat{\rho}({\bf q})$ by Bogoliubov operators leads to an expression for the static
structure factor $S(q)=(u_q - v_q)^2$.   The static structure factor tends to $S(q) \rightarrow
\hbar q / 2 m c$ and vanishes in the long wavelength limit, as required of a zero--temperature
system with finite compressibility \cite{pric54}.

This reveals a remarkable many-body effect.  For free particles,
the matrix element for momentum transfer is always 1, which
reflects the fact that the operator $e^{{\rm i}{\bf q \cdot r}}$
connects an initial state with momentum ${\bf p}$ to a state with
momentum ${\bf p}+\hbar {\bf q}$ with unity overlap.  For an
interacting Bose-Einstein condensate, this overlap vanishes in the
long-wavelength limit. As we have discussed in previous
publications \cite{stam99phon,stamp00leshouches}, this
suppression of momentum transfer is due to an destructive
interference between the two pathways for a condensate to absorb
momentum $\bf q$ and create a quasi-particle: one pathway
annihilates an admixture with momentum $- {\bf q}$, the other
creates an admixture at momentum $+{\bf q}$.

The leading terms in stimulated light scattering do not depend on temperature.  The rates $W_+$
and $W_-$ (Eqs. \ref{eq:net-rate} and \ref{eq:anti-stokes-rate}) are independent of the thermally
excited population of quasiparticles $N_q$ and $N_{-q}$ in the limit of large $n_k,\:n_{k-q}\gg
N_{\pm q}$, i.e. when the scattering is induced by a second laser beam.  For spontaneous scattering
($n_{k-q}=0$), one obtains for the total scattering rate $W$ instead of Eq.
\ref{eq:total-stim-rate}
\begin{equation}
W/N_0=  \frac{2 \pi}{\hbar^2} |C|^2 S(q) n_k \left[(1+N_q) (\delta(\omega - \omega_q^B)+N_{-q}
\delta(\omega +\omega_q^B)) \right]. \label{eq:total-spont-rate}
\end{equation}
Absorption of and bosonic stimulation by thermally excited
quasi-particles become important when the temperature is
comparable to or larger than the quasi-particle energy $\hbar
\omega_q^B$. Eq.\ \ref{eq:total-spont-rate} is proportional to
the temperature dependent dynamic structure factor $S_T({\bf
q},\omega)$\cite{grif93}.

\section{Suppression of impurity collisions}
\label{sec:collisions}

The key difference between the scattering of light and massive particles (or impurities) is their
energy-momentum dispersion relation. The dispersion relation for impurities is $E_{k}=(\hbar
k)^{2}/2M$, whereas for light, $E_{k}=\hbar k c_{l}$ with $c_{l}$ denoting the speed of light.
This difference is responsible for the complete suppression of impurity scattering at low
velocities.  For simplicity, we assume equal mass $M$ for the impurity and condensate atoms.

Energy-momentum conservation, the delta function in Eq.\
\ref{eq:net-rate}, requires $E_k - E_{k-q} = \hbar \omega_q^B$.
For impurity particles, the l.h.s.\ is always less than $v \hbar
q$, where $v=\hbar k/M$ is the initial velocity of the impurities.
Thus, collisions with the condensate are only possible, if this
maximum energy transfer is sufficient to excite a quasi-particle,
i.e.,\ $v>\min(\omega_q^B/q)=v_{L}$, where $v_{L}$ is the Landau
critical velocity~\cite{land41} for superfluidity below which no
dissipation occurs because the density of final states vanishes.

For the excitation spectrum of the condensate (Eq.\
\ref{eq:Bogoliubov-energies}), the Landau velocity is the
Bogoliubov speed of sound $v_{L}=c=\sqrt{\mu/M}$.  Impurity
particles moving below this speed cannot dissipate energy in
collisions.  In contrast, for photons, ${\rm d}E_k/{\rm d}(\hbar
k)=c_{l} \gg c$, and scattering is always possible.

In the perturbative limit (no stimulation by the final
occupation), the total rate of scattering $\Gamma$ is given by
integrating Eq.\ \ref{eq:net-rate} over all possible momentum
transfers.

\begin{eqnarray}
 \Gamma
    & = & \frac{2\pi |C|^{2}}{\hbar^{2}} N_{0}
    \sum\nolimits_{\bf q} {S(q)} \,\,
    \delta \left(\frac{\hbar {\bf k \cdot
    q}}{M}-\frac{\hbar q^{2}}{2 M}- \omega_{q}^{B} \right) \nonumber \\
    &=&  (N_{0}/V)\left(\frac{2\hbar a}{M}\right)^{2}
    \int \!\! dq d\Omega \,\, q^{2} S(q) \,\,
    \delta \left(\frac{\hbar  k q cos\theta}{M}-\frac{\hbar q^{2}}{2 M}-
    \omega_{q}^{B} \right) \nonumber \\
   &=&  2\pi (N_{0}/V)\left(\frac{2\hbar a}{M}\right)^{2} \frac{1}{v}
   \int^{Q}_{0} \!\! dq \,\, q S(q) \nonumber \\
    &=& (N_{0}/V) \: \sigma(\eta) \: v,
    \label{eq:gamma}
\end{eqnarray}
where $\hbar Q= M v (1-1/\eta^{2})$ is the maximum possible
momentum transfer, and $\eta=v/c$ must be larger than 1. We have
used $C=4\pi \hbar^2 a/M V$ for $s$-wave scattering between
condensate and the impurity particles where $a$ denotes the
scattering length and $V$ the condensate volume. The collision
cross section is $\sigma(\eta) = \sigma_0 F(\eta)$ where
$\sigma_0 = 4 \pi a^2$. For $\eta<1$, $F(\eta)=0$ and for
$\eta>1$, $F(\eta)=1-1/\eta^{4}-\log(\eta^{4})/\eta^{2}$.

\begin{figure}
    \epsfxsize=70mm
 \centerline{\epsfbox{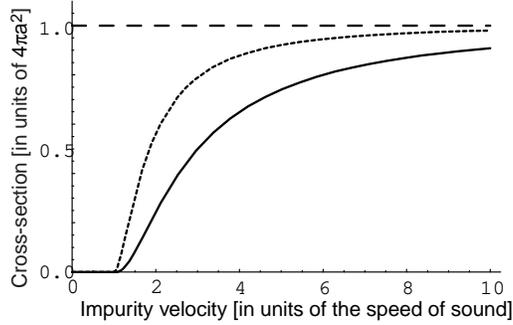}} \caption{Suppression of collisions. Shown is the
suppression factor as a function of the impurity velocity
(normalized by the speed of sound $c$, solid line). The dotted
line represents the suppression due to phase-space restriction
alone (i.e.,\ setting the structure factor $S(q)=1$).}
 \label{coll_fc_theory}
\end{figure}

The suppression factor $F(\eta)$ is determined by two factors: the phase space restriction due to
the delta function in Eq.~\ref{eq:gamma}, and additional suppression at low momentum transfers by
the structure factor of the condensate. For decreasing velocity, the possible scattering angles
become restricted to a forward scattering cone, which shrinks to zero solid angle at the Landau
critical velocity.  A graph of the suppression factor as function of impurity velocity is shown in
Fig.\ \ref{coll_fc_theory}.

Experimentally, the Landau critical velocity can usually only be
observed by moving {\it microscopic} particles through the
superfluid which do not create a macroscopic flow pattern.
Studies of superfluidity with microscopic objects were pursued in
liquid $^{4}$He by dragging negative ions through pressurized
$^{4}$He~\cite{meye61,allu77}, and by scattering $^{3}$He atoms
off superfluid $^{4}$He droplets~\cite{harm99}.

\begin{figure}
        \epsfxsize=90mm
 \centerline{\epsfbox{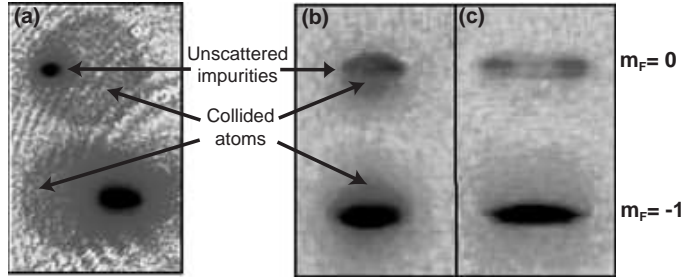}}
\caption{ Observation of elastic collisions between the condensate and impurity atoms. (a)
Impurities traveling at 6 cm/s along the radial axis (to the left in images) were scattered into a
$s$-wave halo. Absorption image after 50 ms of time-of-flight shows the velocity distribution
after collisions between the condensate (bottom) and the outcoupled $m_{F}=0$ atoms (top),
spatially separated by a Stern-Gerlach type magnetic field gradient. The collisional products are
distributed over a sphere in momentum space. The image is 4.5 $\times$ 7.2 mm. (b) Similar image
as (a) shows the collisional products (arrow) for impurity atoms (top) traveling at 7 mm/s along
the condensate axis (upward in image). For this image, $v_{g}/c =2.7$ (see text). Collisions are
visible below the unscattered impurities. (c) Similar image as (b) with $v_{g}/c =1.6$. Collisions
are suppressed. The outcoupled atoms were distorted by mean-field repulsion.  The images are 2.0
$\times$ 4.0~mm.}
    \label{collisions}
\end{figure}

To study the effects of impurities interacting with the
condensate, we created microscopic impurity atoms using a
stimulated Raman process which transferred a small fraction of
the condensate atoms in the $|F=1,m_{F}=-1\rangle$ hyperfine
state into an untrapped hyperfine state $|F=1,m_{F}=0\rangle$
with a well-defined initial velocity~\cite{hagl99}. The initial
velocity could be adjusted between zero and two recoil velocities
by varying the angle between the two Raman beams.  As these
impurities traversed the condensate, they collided with the
stationary condensate, resulting in a redistribution of the
impurity momenta which was detected by a time-of-flight analysis
\cite{chik00} (see Fig.\ \ref{collisions}).

To probe for the suppression of collisions, one has to vary the
impurity velocity around the speed of sound.  For that, we
produced impurity atoms at low velocities (7 mm/s) and varied the
speed of sound by changing the condensate density.  The small
axial velocity imparted by Raman scattering allowed us to
identify products of elastic collisions in time-of-flight images
(Fig.\ \ref{collisions}b, c) since collisions with the stationary
condensate redistributed the impurity atoms toward lower axial
velocities. However, the impurity velocity was predominantly
determined by the gravitational acceleration $g$, which imparted
an average velocity of $v_{g}=\sqrt{2 g z_{c}}$ where $z_{c}$ is
the Thomas-Fermi radius of the condensate in the $z$-direction.
Thus, the effect of superfluidity on impurity scattering depends
primarily on the parameter $\overline{\eta} = v_g / c$ which is
the ratio of the typical impurity velocity $v_g$ to the speed of
sound $c$ at the center of the condensate.

\begin{figure}
    \epsfxsize=60mm
 \centerline{\epsfbox{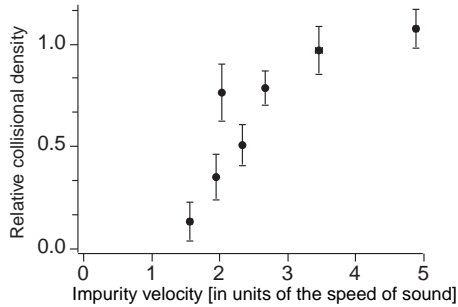}}
\caption{Onset of superfluid suppression of collisions.  Shown is
the observed collisional density normalized to the predicted one
in the limit of high velocities as a function of
$\overline{\eta}=v_{g}/c$, which is a measure of the impurity
velocity in units of the condensate's speed of sound. The error
bars represent the statistical uncertainty. Data is taken from
Ref.\ \protect\cite{chik00} }
 \label{coll_maindata}
\end{figure}

A time-of-flight analysis of impurity scattering for the case of a low-density condensate (small
$c$) and large condensate radius (large $v_g$) is shown in Fig.\ \ref{collisions}b. The effect of
collisions is clearly visible with about 20\% of the atoms scattered to lower axial velocities
(below the unscattered impurities in the image). In contrast, in the case of tight confinement,
the condensate density is higher (larger $c$) and its radius is smaller (smaller $v_g$), and the
collision probability is greatly suppressed due to superfluidity (Figs.\ \ref{collisions}c and
\ref{coll_maindata}).

\begin{figure}
   \epsfxsize=50mm
 \centerline{\epsfbox{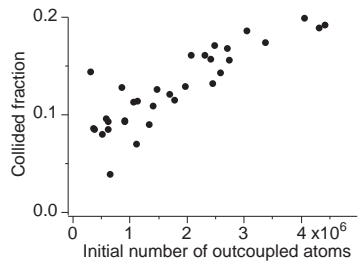}}
\caption{ Collective amplified elastic scattering in a
Bose-Einstein condensate. Shown is the fraction of collided atoms
vs.\ the number of outcoupled atoms. For this data, $v_{g}/c$=4.9
and the chemical potential was 1.8 kHz. Figure is taken from
Ref.\ \protect\cite{chik00} }
 \label{coll_enhance}
\end{figure}

The bosonic stimulation factor in Eq.\ \ref{eq:net-rate} is
relevant if the final states are populated, either by scattering
or thermally.  We observed that the fraction of collided atoms
increased with the number of outcoupled impurities
(Fig.~\ref{coll_enhance}). For a large outcoupled fraction,
population $n_{k-q}$ and $N_q$ is built up in the final states
and stimulates further scattering. This collisional amplification
is not directional, and is similar to the recently observed
optical omnidirectional superfluoresence~\cite{lvov99}.

Gain of momentum and thus transfer of energy from the condensate
to the impurity atoms is impossible at zero temperature, but may
happen at finite temperature due to the presence of thermal
excitations (the $N_{-q}$ term in Eq.\
\ref{eq:anti-stokes-rate})\footnote{We are grateful to S.
Stringari for pointing out the importance of finite-temperature
effects.}. Thus finite temperature enhances the elastic cross
section by two effects: Absorption of quasi-particles
(anti-Stokes process) and stimulation of momentum transfer by the
final state population (Stokes process).

\begin{figure}
    \epsfxsize=70mm
\centerline{\epsfbox{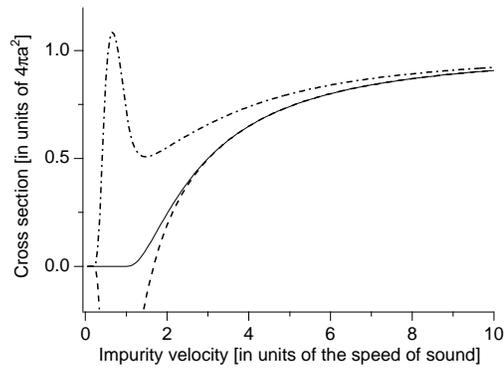}}
\caption{Temperature dependent cross-section vs.\ impurity
velocity. Shown is the cross-section at zero temperature (solid
line) and at a finite temperature $kT=\mu$ which is typical for
our experimental conditions (dash-dotted line). The finite
temperature cross-section includes collisions involving thermally
occupied quasi-particles where the impurities lose or gain
energy. In the experiment, we measured the number of impurities
which lost its energy minus the number which gained energy. Thus,
the experimental measured cross-sections (Fig.\
\protect\ref{coll_maindata}) should be compared to $\sigma_{\rm
coll,loss} -\sigma_{\rm coll,gain}$ (dashed line).}
\label{finitetemp}
\end{figure}

Fig.\ \ref{finitetemp} shows the dramatic variation of the
elastic scattering cross section with temperature. However, the
finite temperature did not affect our data in a major way:  Due to
gravitational acceleration we couldn't probe the velocity regime
well below the Landau critical velocity where only thermally
assisted collisions are possible.  Furthermore, when we counted
the number of collided atoms we had to use a background
subtraction method where we subtracted the small signal of the
energy gain collisions from the energy loss collisions (see Ref.\
\cite{chik00} for details), thus cancelling most of the
finite-temperature effects.

\section{Suppression of dissipation for a moving macroscopic object}
\label{sec:macroscopic_flow} So far, we have discussed the suppression and enhancement of
microscopic processes (light scattering and impurity collisions).  The suppression of dissipation
is even more dramatic on the macroscopic scale.  The flow of liquid $^4$He and the motion of
macroscopic objects through it are frictionless below a critical velocity \cite{nozi90}.
Recently, we have explored such frictionless flow in a gaseous BEC \cite{rama99,onof00sup}.

The microscopic and macroscopic cases bear many parallels.  The
onset of scattering or dissipation has two requirements:  one
needs final states which conserve energy and momentum, and an
overlap matrix element which populates these states.  In the case
of macroscopic flow, the first requirement leads to a critical
velocity for vortex creation and the second requirement addresses
the nucleation process of vortices.

The Landau criterion for superfluidity shows that excitations
with momentum $p$ and energy $E(p)$ are only possible when the
relative velocity between the fluid and the walls or a
macroscopic object exceeds the Landau critical velocity $v_L$
which is given by $v_L = \min(E(p)/p)$ (see e.g.\
\cite{huan87,nozi90}. A similar criterion applied to vortex
formation yields
\begin{equation}
v_{c}  = \frac{E_{\rm vortex}}{I_{\rm vortex}} \sim \frac{\hbar
}{{M d}}\ln \left( {\frac{d}{\xi }} \right) \label{eq:feyn}
\end{equation}
where $I = \int{P d^3 r}$ is the integrated momentum of the
vortex ring or line pair, $E$ is its total energy, $d$ is the
dimension of the container, and $\xi$ the core radius of a vortex
which in the case of dilute gases is  the healing length $\xi =
1/\sqrt{8 \pi n a}$. Ref.\ \cite{nozi90} derived Eq.\
\ref{eq:feyn} for vortex rings with a maximum radius $d$, Ref.\
\cite{cres00} looked at pairs of line vortices at distance $d$.
Feynman \cite{feyn55} found a similar result for superflow
through a channel of diameter $d$.

A similar result is obtained for a Bose condensed system placed under uniform rotation with angular
velocity $\Omega$.  A vortex becomes energetically allowed when its energy $E'$ in the rotating
frame drops to zero,
\begin{equation}
E' = E - \Omega L = 0
\end{equation}
where $E$ and $L$ are the energy and angular momentum in the
laboratory frame.  This defines a critical {\em angular} velocity
below which a vortex cannot be sustained due to conservation of
angular momentum and energy~\cite{lund97}:

\begin{equation}
\Omega_c = \frac{E_{\rm vortex}}{L_{\rm vortex}} \sim
\frac{\hbar}{M d^2}\ln{\left(\frac{d}{\xi}\right)}.
\label{eq:vortex}
\end{equation}

The critical velocity at the wall of the rotating container, $v_{c}=d \Omega_c$, agrees with Eq.\
\ref{eq:feyn}.

However, Eqs.\ \ref{eq:feyn} and \ref{eq:vortex} only reflect the
energy and momentum required to generate vortices, and do not
take into account the nucleation process.  If the scattering
particle is macroscopic in size, the coupling is between the
ground state and a state containing a vortex. Populating such a
state requires nucleation of the vortex by the perturbing
potential, which usually does not occur until higher velocities
are reached than those predicted by Eqs.\ \ref{eq:feyn} and
\ref{eq:vortex}. The other option, the formation of the vortex by
macroscopic quantum tunneling between the two states is an
extremely slow process. In recent experiments in which a Bose
condensate was placed in a rotating potential, the critical
angular velocity for the formation of a single vortex was
observed to be 1.6 times higher than the value given by Eq.\
\ref{eq:vortex} \cite{madi00}.  This discrepancy may be due to a
nucleation barrier associated with the excitation of surface
modes, as some authors have recently suggested
\cite{fede99A,isos99vort}.

To study frictionless flow in a Bose-condensate, we focused a
blue-detuned 514 nm Argon laser beam onto the sodium condensate,
which repelled atoms from the focus. The laser beam was scanned
back and forth along the axial direction of the condensate,
creating a moving ``hole'' that simulated a macroscopic object.
Rapid sequence phase-contrast imaging allowed us to directly
measure the flow pattern of the superfluid around the moving
laser beam.

For a weakly interacting Bose-condensed gas at density $n({\bf r})$ and chemical potential $\mu$,
pressure is identical to the mean-field energy density $ P = \mu({\bf r}) n({\bf r})/2$
\cite{dalf99rmp}. A drag force arises due to the pressure difference across the moving object. The
chemical potential is given by $\mu({\bf r},t)=g n({\bf r},t)$, where $g = 4 \pi \hbar^2 a /M$ is
the strength of two-body interactions. The drag force $F$ is given by
\begin{equation} F \simeq  g S n {\bf \Delta} n=S \mu \Delta \mu/g
\label{eq:drag_force}
\end{equation}
where $\Delta n$ and $\Delta \mu$ are the differences in density and chemical potential across the
stirring object, and $S$ the surface area the macroscopic object presents to the condensate.

\begin{figure}[htbf]
\epsfxsize=70mm
\centerline{\epsfbox{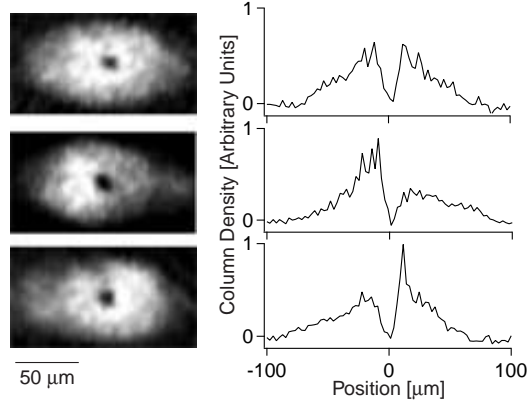}}\vspace{0.1cm}
 \caption{Pressure difference across a laser beam
moving through a condensate. On the left side {\em in situ} phase
contrast images of the condensate are shown, strobed at each
stirring half period: beam at rest (top); beam moving to the left
(middle) and to the right (bottom). The profiles on the right are
horizontal cuts through the center of the images. The stirring
velocity  and the maximum sound velocity were 3.0 mm/s and 6.5
mm/s respectively. Figure is taken from Ref.\ \protect
\cite{onof00sup}. } \label{fig:strobed}
\end{figure}
If the laser beam is stationary, or moves slowly enough to
preserve the superfluid state of the condensate, there will be no
gradient in the chemical potential across the laser focus, and
therefore zero force according to Eq.\ \ref{eq:drag_force}. A
drag force between the moving beam and the condensate is
indicated by an instantaneous density distribution $n(\bf{r},t)$
that is distorted asymmetrically with respect to the laser beam.
Fig.\ \ref{fig:strobed} shows phase contrast images strobed at
half the stirring period (where the laser beam is in the center
of the condensate).  A bow wave and stern wave form in front of
and behind the moving laser beam, respectively.  We define the
asymmetry $A$ as the relative difference between the peak column
densities in front ($\tilde{n}^f$) and behind ($\tilde{n}^b$) the
laser beam $A = 2(\tilde{n}^{f}-\tilde{n}^{b})/(\tilde{n}^{f}+
\tilde{n}^{b})$. The asymmetry $A$ is proportional to the drag
force $F$.

\begin{figure}[htbf]
\epsfxsize=70mm \centerline{\epsfbox{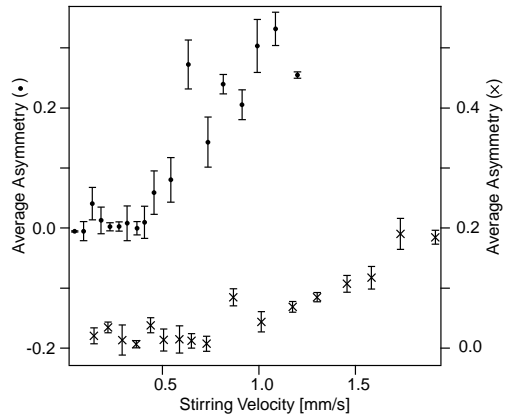}}
\vspace{0.1cm} \caption{Density dependence of the critical
velocity. The onset of the drag force is shown for two different
condensate densities, corresponding to maximum sound velocities
of 4.8 mm/s ($\bullet$, left axis) and 7.0 mm/s ({\bf $\times$},
right axis). The stirring amplitudes are 29 $\mu$m and 58 $\mu$m,
respectively.  The two vertical axes are offset for clarity. The
bars represent statistical errors. Figure is taken from Ref.\
\protect \cite{onof00sup}. }\label{fig:asym_2}
\end{figure}

In Fig.\ \ref{fig:asym_2} we show measurements of the asymmetry
for two maximum densities $n_0$ of $9 \times 10^{13}$ and $1.9
\times 10^{14}$ cm${}^{-3}$.  In each data set there is a
threshold velocity $v_c$ below which the drag force is
negligible, and this threshold increases at higher density.  Its
value is close to $0.1\ c$ for both data sets, where $c$ is the
sound velocity. Above this critical velocity, the drag force
increases monotonically, with a larger slope at low density.  In
addition, the heating rate due to friction against the laser beam
was directly measured through time of flight absorption imaging
\cite{rama99,onof00sup,rama00jltp} and found to be in good
agreement with the value ${\bf F} \cdot {\bf v}$, with ${\bf F}$
estimated from Eq.\ \ref{eq:drag_force}.

The observed critical velocity may be related to the formation of
vortices. An estimate based on Eq.\ \ref{eq:feyn} for typical
experimental parameters in sodium ($d = 10\ \mu$m, peak density
$n_0 = 1.5 \times 10^{14}$ cm$^{-3}$, $a = 2.75$ nm) yields $v_c
\simeq 1.0$ mm/s, close to the experimental observations.
However, Eq.\ \ref{eq:feyn} depends only weakly on the speed of
sound, through the logarithmic dependence on the healing length
$\xi$. In contrast, our measurements show an approximate
proportionality to the sound velocity\cite{onof00sup}, suggesting
that vortex nucleation, determines the onset of dissipation.

Time-dependent simulations of the Gross-Pitaevskii equation show
the formation of vortex line pairs, above a critical velocity
which is close to the observed value \cite{jack00}.  Several
authors have emphasized the role of locally supersonic flow
around the laser beam in the nucleation of vortices.
\cite{jack00,fris92,jack99}.  In one theoretical model
\cite{fris92}, the vortices are emitted periodically at a rate
that increases with velocity, and reduce the pressure gradient
across the object.  The predicted heating rate
\cite{fris92,wini99} is in rough agreement with the data.
Moreover, this model also predicts that the slope of the
asymmetry should increase at lower density, in accord with our
observations.

\section{Amplification of light and atoms}
\label{sec:amp_light_atoms}

Spontaneous light scattering can be stimulated when the atomic recoil state is already populated
(the $N_q$ term in Eq.\ \ref{eq:net-rate}). We have explored this process in our studies of
superradiance \cite{inou99super}, phase-coherent atom amplification \cite{inou99mwa} and optical
amplification \cite{inou00slow}.

In all these experiments, the condensate was illuminated with a laser beam (mode $k$, also called
``dressing beam'').  A condensate atom scatters a photon from the laser beam into another mode and
receives the corresponding recoil momentum and energy.  Injection of atoms or light turns this
{\it spontaneous} process into a {\it stimulated} process and realizes an amplifier for either
atoms or light. If atoms are injected, they form a matter wave grating (an interference pattern
with the condensate at rest) and this grating diffracts light. The diffraction transfers recoil
momentum and energy to the atoms, which results in a growth of the grating and therefore the
number of atoms in the recoil mode --- this is the intuitive picture for atom gain.

Eq.\ \ref{eq:net-rate} describes gain for atoms. In the limit of an empty mode for the scattered
light ($n_{k-q}=0$), one obtains
\begin{equation}
\label{eq:gain-rate}
  W_+ =\frac{2 \pi}{\hbar} |C|^2 N_0 n_k (N_q +1) \delta(E_k-E_{k-q} - \hbar \omega_q^B)
\end{equation}
For the high  momentum transfers considered here (on the order of
the photon recoil momentum), $S(q)=1$. Each scattering event
which transfers momentum ${\bf q}$ to the condensate, generates a
recoiling atom---therefore the scattering rate $W_+$ integrated
over all final states for the scattered photon gives the growth
rate $\dot{N_{q}}$ for the recoiling atoms:
\begin{equation}
\dot{N_{q}}=(G_{q}-\Gamma_{2,q}) N_{q}
\label{eq:gain-equation}
\end{equation}
with the gain coefficient
\begin{equation}
G_{q}=R\,N_{0}\,\frac{\sin^{2}{\theta_{q}}}{8 \pi/3}\Omega_{q}.
\label{eq:gain-coeff}
\end{equation}
Here $R$ is the rate for single-atom Rayleigh scattering which is proportional to the pump light
intensity, $N_0$ the number of atoms in the condensate at rest, $\theta_{q}$ the angle between the
polarization  of the incident light and the direction of the scattered light, and $\Omega _{q}$ the
phase-matching solid angle for scattering into mode $q$.  In addition, a loss term $\Gamma_{2,q}$
was included which describes the decoherence rate of the matter-wave grating and determines the
threshold for exponential growth.  It represents the linewidth of the two-photon process
generating recoil atoms in mode $q$.

\begin{figure}
    \epsfxsize=65mm
\centerline{\epsfbox{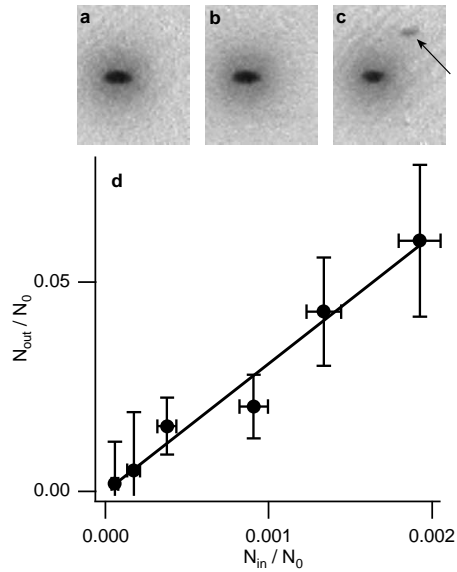}}
    \caption{Input--output characteristic of the matter-wave amplifier.  ({\bf a-c}) Typical
time-of-flight absorption images demonstrating matter wave
amplification. The output of the seeded amplifier ({\bf c}) is
clearly visible, whereas no recoiling atoms are discernible in
the case without amplification ({\bf a}) or amplification without
the input ({\bf b}). The size of the images is 2.8~mm $\times$
2.3~mm. ({\bf d}) Output of the amplifier as a function of the
number of atoms at the input.  A straight line fit shows a number
gain of 30.  Reprinted by permission from Nature, Ref.~\protect
\cite{inou99mwa}, copyright 1999 Macmillan Magazines Ltd.
    \label{fig:MWAgain}}
\end{figure}

\begin{figure}
    \epsfxsize=70mm
\centerline{\epsfbox{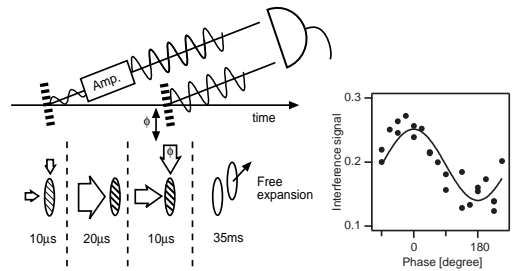}}
    \caption{ Experimental scheme for observing phase coherent matter wave amplification. A
small-amplitude matter wave was split off the condensate by
applying a pulse of two off-resonant laser beams (Bragg pulse).
This input matter wave was amplified by passing it through the
condensate pumped by a laser beam. The coherence of the amplified
wave was verified by observing its interference with a reference
matter wave, which was produced by applying a second (reference)
Bragg pulse to the condensate.  The interference signal was
observed after 35~ms of ballistic expansion. The fringes on the
right side show the interference between the amplified input and
the reference matter wave.  Reprinted by permission from Nature,
Ref.~\protect \cite{inou99mwa}, copyright 1999 Macmillan
Magazines Ltd.
    \label{fig:MWAsetup}}
\end{figure}

Fig.~\ref{fig:MWAgain} shows the input-output characteristics of the amplifier. The number of
input atoms was below the detection limit of our absorption imaging (Fig.~\ref{fig:MWAgain}a). The
amplification pulse alone, although above the threshold for superradiance \cite{inou99super}, did
not generate a discernible signal of atoms in the recoil mode (Fig.~\ref{fig:MWAgain}b). When the
weak input matter wave was added, the amplified signal was clearly visible
(Fig.~\ref{fig:MWAgain}c).  The gain was controlled by the intensity of the pump pulse (see
Eq.~\ref{eq:gain-coeff}) and typically varied between 10 and 100. Fig.~\ref{fig:MWAgain}d shows
the observed linear relationship between the atom numbers in the input and the amplified output
with a number gain of 30.

The phase of the amplified matter wave was determined with an interferometric technique.  For this,
a reference matter wave was split off the condensate in the same way as the first (input) wave
(see Fig.\  \ref{fig:MWAsetup}). The phase of the reference matter wave was scanned by shifting the
phase of the radio-frequency signal that drove the acousto-optic modulator generating the axial
Bragg beam. We then observed the interference between the reference and the amplified matter waves
by measuring the number of atoms in the recoil mode.

The atom amplification is described by the Hamiltonian (Eq.\
\ref{eq:hamil}) as a four-wave mixing process between two
electromagnetic fields and two Schrödinger fields.  The symmetry
between light and atoms indicates that a dressed condensate
should not only amplify injected atoms, but also injected light.
In Ref.\ \cite{stamp00leshouches} we have discussed that matter
wave gain and optical gain emerge as two limiting cases,
depending on whether the atomic population $N_{q}$ or occupation
of the optical mode $n_{k-q}$ dominates the bosonic stimulation
term in Eq.\ \ref{eq:net-rate2}.  In our experiments, the
recoiling atoms move ten orders of magnitude more slowly than the
scattered light.  Therefore, we always have $N_{q} \gg n_{k-q}$
and $n_{k-q}<1$ and optical stimulation should play no role.
Still, as we want to discuss now, the dressed condensate can act
as an amplifier for light.

\begin{figure}[htbf]
\epsfxsize=69mm
\centerline{\epsfbox{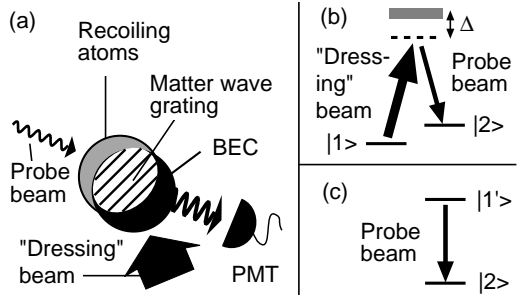}}\vspace{0.5cm}
\caption{Amplification of light and atoms by off-resonant light
scattering. (a) The fundamental process is the absorption of a
photon from the ``dressing'' beam by an atom in the condensate
(state $|1 \rangle$), which is transferred to a recoil state
(state $|2 \rangle$) by emitting a photon into the probe field.
The intensity in the probe light field was monitored by a
photomultiplier. (b) The two-photon Raman-type transition between
two motional states ($|1\rangle, |2\rangle$) gives rise to a
narrow resonance. (c) The dressed condensate is the upper state
($|1^{\prime}\rangle$) of a two-level system, and decays to the
lower state (recoil state of atoms, $|2\rangle$) by emitting a
photon. Figure is taken from Ref.\ \protect\cite{inou00slow}}
\label{fig:setup-light-amp}
\end{figure}

The physical picture behind the optical gain is as follows: If a very weak probe beam is injected
into the dressed condensate, it acts together with the dressing beam as a pair of Bragg beams and
creates recoiling atoms.  Those recoiling atoms move out of the condensate (or decohere) on a time
scale $\Gamma_2^{-1}$ which is the inverse of the linewidth of the Bragg transition.  In steady
state, the number of recoiling atoms $N_q$ in the volume of the condensate is proportional to the
intensity of the probe light.  Those recoiling atoms interfere with the condensate at rest and
form a diffraction grating which diffracts the dressing beam into the path of the probe light
resulting in amplification of the probe light (Fig.\ \ref{fig:setup-light-amp}).

An expression of the gain can be derived in analogy to a fully
inverted two-level system with dipole coupling which would have a
gain cross section of $6\pi\lambdabar^{2}$ for radiation with
wavelength $\lambda (= 2\pi\lambdabar)$. For the Raman-type
system in Fig.\ \ref{fig:setup-light-amp}b, the gain is reduced
by the excited state fraction, $R / \Gamma_1$ (where $R$ is the
Rayleigh scattering rate for the dressing beam and $\Gamma_1$ is
the linewidth of the single-photon atomic resonance) and
increased by $\Gamma_1/\Gamma_{2}$, the ratio of the linewidths
of the single-photon and two-photon Bragg resonances. Thus the
expected cross-section for gain is
\begin{equation}
\sigma_{\rm gain}=6\pi\lambdabar^{2} \frac{R}{\Gamma_{2}}.
\label{eq:twophotongain}
\end{equation}

\begin{figure}[htbf]
\epsfxsize=69mm \centerline{\epsfbox{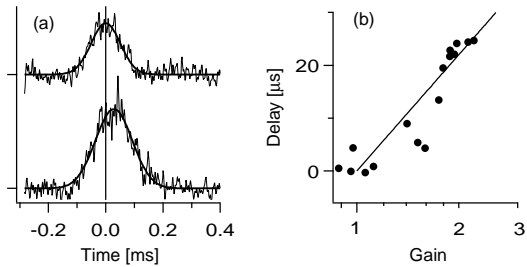}}
\vspace{0.5cm} \caption{Pulse delay due to light amplification.
(a) About 20 $\mu$s delay was observed when a Gaussian pulse of
about 140 $\mu$s width and 0.11 mW/cm$^{2}$ peak intensity was
sent through the dressed condensate (bottom trace). The top trace
is a reference taken without the dressed condensate. Solid curves
are Gaussian fits to guide the eyes. (b) The observed delay was
proportional to the logarithm of the observed gain. Figure is
taken from Ref.\ \protect\cite{inou00slow}} \label{fig:gain-delay}
\end{figure}

We observed quasi-steady state gain by a factor of up to three.
This optical gain has a narrow bandwidth due to the long
coherence time of a condensate.  The gain represents the
imaginary part of the complex index of refraction.  A sharp peak
in the gain implies a steep dispersive shape for the real part of
the index of refraction $n(\omega)$.  This resulted in an
extremely slow group velocity for the amplified light.  Fig.\
\ref{fig:gain-delay} shows that light pulses were delayed by
about 20 $\mu$s across the 20 $\mu$m wide condensate
corresponding to a group velocity of 1 m/s.  This is one order of
magnitude slower than any value reported previously (see Ref.
\cite{hau99} and references therein).

The (amplitude) gain $g$ for the probe light is related to the matter wave gain $G$
\begin{equation}
g=1+ \frac{n_{0}\sigma_{\rm gain} l}{2}=1+\frac{G}{\Gamma_2}
\label{eq:gain1}
\end{equation}
where $n_0$ is the condensate density and $l$ its length.
However, when the gain $G$ is above the threshold for
superradiance, $G>\Gamma_2$ (Eq.\ \ref{eq:gain-equation}) the
optical gain should diverge: a single recoiling atom created by
the probe light and dressing light is exponentially amplified and
creates a huge matter wave grating which will diffract the
dressing light into the probe light path, thus amplifying the
probe light by a large factor. In order to describe this
non-linear feedback, one has to use coupled equations for optical
gain and atom gain \cite{inou00slow}.  As a result, one has to
replace the loss rate $\Gamma_2$ in Eq.\ \ref{eq:gain1} by the
dynamic rate $\Gamma_2-G$ (see Eq.\ \ref{eq:gain-equation})
\begin{equation}
g=1+\frac{G} {\Gamma_2-G}=1+ \frac{n_{0}\sigma_{\rm gain} l}{2}
\frac{\Gamma_2}{\Gamma_2-G}
\label{eq:gain2}
\end{equation}
which agrees with Eq.\ (\ref{eq:gain1}) in the low-intensity
limit.  By raising the gain over the threshold we could map out
the transition from single-atom gain to collective
gain~\cite{inou00slow}. The expansion $g=1+ (G/\Gamma_2) +
(G/\Gamma_2)^2 + \ldots $ shows the transition from (linear)
single-atom gain to (non-linear) collective gain and illustrates
that the dressed condensate is a clean model system for
discussing linear and non-linear behavior, optical and
atom-optical properties and their interplay.

\section{Discussion}

This paper has summarized our recent experiments on Bose-Einstein
condensation with the unifying theme of suppression and
enhancement. Suppression of scattering or dissipation can arise
for two different reasons. The phonon and vortex nature of the
collective excitations together with energy and momentum
conservation allow dissipation only above a critical velocity. In
addition, one has to consider the dynamics of the excitation
process. For microscopic particles, this is reflected in the
matrix element $S(q)$ which characterizes how easily can the
condensate absorb momentum. For macroscopic motion, it is
reflected in a critical velocity for vortex nucleation.
Scattering processes are also enhanced by the population in the
final states (bosonic stimulation). Optical stimulation was used
in Bragg scattering, and matter wave stimulation led to
superradiance and matter wave amplification.

In closing, let us note that there are some subtleties which go
beyond the simple picture using rate equations and occupation
numbers.  A condensate in its ground state is in a coherent
superposition state of the zero-momentum state with correlated
pairs with momenta $\pm {\bf q}$ (the quantum depletion). The
population in the quantum depletion can cause bosonic stimulation
of spontaneous emission \cite{gorl00spont}. However, for a
scattering situation, there are two bosonically enhanced pathways
which destructively interfere (causing $S(q) < 1$, Sect.\
\ref{sec:Bragg_spect}).  Therefore, the concept of bosonic
stimulation can be applied to the quantum depletion, but with
caution \cite{gorl00spont}.

Finally, the matter wave amplification described in Sect.\
\ref{sec:amp_light_atoms} can be regarded as the
self-amplification of a density modulation (caused by matter wave
interference between the condensate and the amplified recoiling
atoms). A similar amplification can happen for a density
modulation in a fermionic gas \cite{kett00fermi}, but, the
coherence time of such a density modulation is generally much
shorter than in a condensate.  This example shows that a more
general description of stimulation and amplification has to
address the symmetry and coherence of the prepared state. Bosonic
quantum-degeneracy is sufficient, but not necessary for
stimulated scattering \cite{kett00fermi}.

We are grateful to Axel G\"{o}rlitz for helpful discussions. This work was supported by NSF, ONR,
ARO, NASA, and the David and Lucile Packard Foundation.

\end{document}